\documentclass[10pt, pre, twocolumn, showpacs, aps, floatfix]{revtex4-1}
\usepackage{amsmath,amssymb}
\usepackage{bm}
\usepackage[dvipdfmx]{graphicx,color}
\usepackage{multirow}
\usepackage{tabularx}
\usepackage{ulem}
\newcommand{\diff}{\mathrm{d}}

\begin{document}
\title{Critical exponents in coupled phase-oscillator models on small-world networks}
\author{Ryosuke Yoneda, Kenji Harada, and Yoshiyuki Y. Yamaguchi}
\affiliation{Graduate School of Informatics, Kyoto University, 606-8501 Kyoto, Japan}
\begin{abstract}
  A coupled phase-oscillator model consists of phase-oscillators,
  each of which has the natural frequency obeying a probability distribution
  and couples with other oscillators through a given periodic coupling function.
  This type of model is widely studied since it describes the synchronization
  transition, which emerges between the non-synchronized state and
  partially synchronized states.
  The synchronization transition is characterized by several critical exponents,
  and we focus on the critical exponent defined
  by coupling strength dependence of the order parameter
  for revealing universality classes.
  In a typical interaction represented by the perfect graph,
  an infinite number of universality classes is yielded
  by dependency on the natural frequency distribution and the coupling function.
  Since the synchronization transition is also observed in a model
  on a small-world network, whose number of links is proportional
  to the number of oscillators,
  a natural question is whether the infinite number of universality classes
  remains in small world networks irrespective of the order of links.
  Our numerical results suggest that the number of universality class
  is reduced to one and the critical exponent
  is shared in the considered models
  having coupling functions up to the second harmonics
  with unimodal and symmetric natural frequency distributions.
  
\end{abstract}
\maketitle

\section{Introduction}
Ever since Huygens found that two pendulum clocks hanging on a wall swung in the opposite direction from each other,
many illustrations of synchronization have been established in various fields of nature,
such as frog choruses~\cite{aihara2014}, flashing of fireflies~\cite{smith1935,buck1968},
metronomes~\cite{pantaleone2002}, and circadian rhythms~\cite{winfree1967}.
It is natural to try to understand synchronization theoretically,
and a coupled phase-oscillator model is one of successful models to describe synchronization~\cite{kuramoto2019}.
This model consists of many coupled oscillators,
and the coupling is expressed by a periodic coupling function.
Each oscillator has the so-called natural frequency, randomly drawn from a natural frequency distribution.
When the coupling strength $K$ increases,
the oscillators exhibit the synchronization transition
from the non-synchronized state to (partially) synchronized states.
The synchronization transition is continuous or discontinuous,
depending on the natural frequency distribution and the coupling function~\cite{kuramoto1975,strogatz2000,chiba2015,daido2015,basnarkov2007,pazo2005,daido1990,crawford1995,chiba2011,pikovsky2013,komarov2014}.

The critical phenomena have been extensively studied in statistical mechanics.
One of their remarkable features is the existence of universality classes;
the systems in a universality class share the critical exponents defined around the critical point $K=K_{\mathrm{c}}$ of a continuous transition.
One of the critical exponents is $\beta$,
defined by $r\sim(K-K_{\mathrm{c}})^{\beta}$, where $r$ is the order parameter.
Thus, it is natural to ask the universality classes in the coupled phase-oscillator models
through values of the critical exponent $\beta$.

For the all-to-all and uniform coupling, extended researches have revealed that the value of $\beta$
depends on the coupling function and the natural frequency distribution~
\cite{kuramoto1975,strogatz2000,chiba2015,daido2015,basnarkov2007,pazo2005,daido1990,crawford1995,chiba2011,pikovsky2013,komarov2014}.
For simplicity, we focus on coupling functions which have two harmonics at most,
and review values of the critical exponent $\beta$ for the following three cases:
(i) the second harmonics is absent,
(ii) the second harmonics has the opposite sign with the leading harmonics,
and (iii) the second harmonics has the same sign with the leading harmonics.
We assume that the natural frequency distribution is unimodal and symmetric,
and that the second-leading term of its Maclaurin expansion is of the order $2n$, where $n\in\mathbb{N}$.
A Gaussian distribution and a Lorentzian distribution have $n=1$ for instance.

In the case (i) and (ii), the model shows a continuous transition,
whereas in the case (iii), a discontinuous transition occurs \cite{chiba2011}, 
hence we cannot define the critical exponent $\beta$.
In the case (i), the model becomes the Kuramoto model~\cite{kuramoto1975}, a paradigmatic coupled phase-oscillator model.
Several researches have pointed out that the critical exponent $\beta=1/(2n)$~\cite{kuramoto1975,strogatz2000,chiba2015,daido2015}.
This $n$ dependence is a strong feature of the Kuramoto model
and gives a sharp contrast with the case (ii).
In the case (ii), the critical exponent $\beta$ becomes $1$ for $n=1$~\cite{crawford1995,chiba2011,pikovsky2013,komarov2014},
and this value is suggested to be universal irrespective of $n\in\mathbb{N}$~\cite{chiba2011}.

Apart from the all-to-all coupling,
couplings represented by complex networks are of interests
like random graphs, scale-free networks, and small-world networks \cite{dorogovtsev2008}.
In particular,
we focus on the small-world network
because it is ubiquitous in the real world \cite{watts1998},
and it is a notable network for the synchronization.
The synchronization transition appears
with the critical exponent $\beta=1/2$ in small-world networks
even if they are very close to the one-dimensional lattice \cite{hong2002},
while the one-dimensional lattice hardly shows the synchronization
\cite{sakaguchi1987,daido1988,hong2002}.
The previous research \cite{hong2002} however
lacks to consider universality
since it has treated only the case (i) with $n=1$,
whereas other universality classes might be hidden in other cases
as mentioned above.
In this paper, we numerically study the synchronization transitions
on small-world networks in all the cases (i), (ii), and (iii)
with varying the value of $n$.
Our results suggest that the critical exponent is
universally $\beta=1/2$ for any $n\in\mathbb{N}$
in the cases (i) and (ii), where the transition is continuous,
while discontinuity in the case (iii) is inherited.

This paper is organized as follows.
In Sec.~\ref{sec:model}, we briefly introduce the small-world network and coupled phase-oscillator models on it.
We also introduce a family of the natural frequency distributions, whose second-leading term is of the order $2n$.
In Sec.~\ref{sec:nu-sim}, we show the finite-size scaling
to calculate the critical exponent $\beta$.
A similarity between systems on the small-world network
and noisy systems is discussed in Sec.~\ref{sec:SW-noise}.
Finally, in Sec.~\ref{sec:conclusion}, we summarize this paper and note some future works.

\section{Coupled phase-oscillator models on small-world networks}
\label{sec:model}

A coupled phase-oscillator model is defined by
\begin{equation}
  \begin{split}
    &\frac{\diff\theta_{i}}{\diff t}=\omega_{i}+\frac{K}{2k}\sum_{j\in\Lambda_{i}}f_{a}(\theta_{j}-\theta_{i}),\\
    &f_{a}(\theta)=\sin\theta+a\sin2\theta,
    \label{eq:sw-model}
  \end{split}
\end{equation}
for $i=1,\cdots,N$.
$\theta_{i}$ and $\omega_{i}$ are the phase and the natural frequency of the $i$th oscillator respectively,
and $\omega_{i}$ is randomly drawn from a natural frequency distribution $g(\omega)$.
$K>0$ is a coupling constant,
describing how strong the coupling between oscillators are.
The index set $\Lambda_{i}$ contains the indexes of oscillators
connecting to the $i$th oscillator,
and it determines the network of couplings.
For instance, the all-to-all coupling gives $\Lambda_{i}=\{1,\cdots,N\}$,
and the nearest neighbor coupling on the one-dimensional lattice
gives $\Lambda_{i}=\{i-1,i+1\}$.

\begin{figure}
  \begin{center}
    \includegraphics[width=8cm]{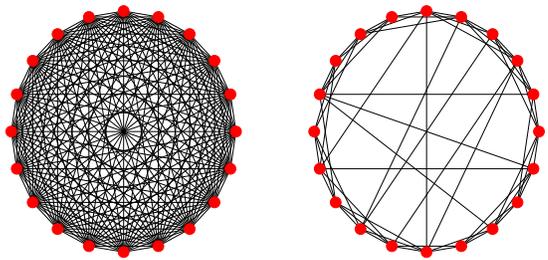}
    \caption{Comparison between the all-to-all network (left) and a small-world network (right) with 20 nodes. 
    The small-world network is constructed from the $k$-nearest neighbour lattice ($k=3$) with the rewriting probability $p=0.2$.}
    \label{fig:all-sw}
  \end{center}
\end{figure}

The coupling network represented by $\{\Lambda_{i}\}_{i=1}^{N}$
is arbitrarily chosen. In this paper, we are interested in
the small-world network,
which possesses the property of a small diameter and a large clustering
coefficient despite its sparsity.
The small-world network can be seen in various fields of the real world,
such as human relationships, World Wide Web, citations of scientific papers,
and so on.
In 1998, Watts and Strogatz proposed a breakthrough network model showing the property of small-world network,
which is created in the following algorithm~\cite{watts1998}.
We first make a periodic $k$-nearest neighbor network with $N$ nodes, which results in $kN$ links.
Then we rewire each link with probability $p$,
keeping in mind that we do not allow self-loops or link duplications.
Moreover, we use only connected small-world networks:
if a generated network is disconnected,
we discard it and generate another one until connected one is created.
See Fig.~\ref{fig:all-sw} for a comparison between the all-to-all network
and a small-world network.
In this paper,
we use the Watts--Strogatz small-world network with
$k=3$ and $p=0.2$, following the previous research \cite{hong2002}
which shows emergence of the synchronization transition
on a small-world network.

\begin{figure}
  \begin{center}
    \includegraphics[width=8cm]{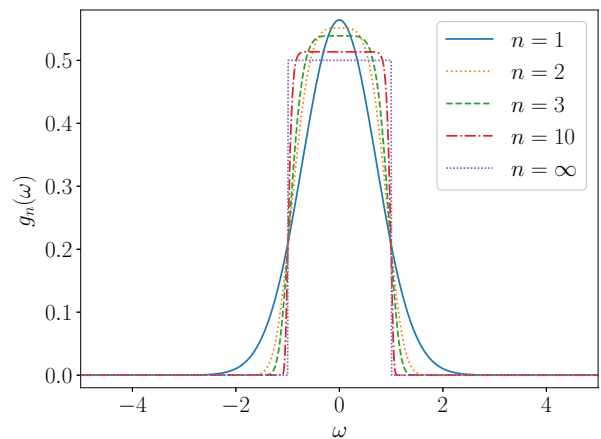}
    \caption{Graphs of $g_{n}(\omega)$ with $n=1,2,3,10,$ and $\infty$,
    where we set $\Delta=1$.
    $g_{n}(\omega)$ converges to $g_{\infty}(\omega)$
    in the limit $n\to\infty$.
    }
    \label{fig:g_n}
  \end{center}
\end{figure}

As the natural frequency distribution $g(\omega)$,
we introduce a family of distributions
parametrized by a natural number $n\in\mathbb{N}$,
\begin{align}
  g_{n}(\omega)=\frac{n}{\Gamma(1/(2n))\Delta}e^{-(\omega/\Delta)^{2n}},
  \label{eq:g_n}
\end{align}
where $\Gamma(z)=\int_{0}^{\infty}t^{z-1}e^{-t}\diff t$ is the Gamma function defined on $\Re(z)>0$.
Here, $\Delta>0$ is a parameter describing the width of the distribution. 
We note that $n=1$ gives the Gaussian distribution.
The distribution $g_{n}(\omega)$
is unimodal and symmetric with respect to $\omega=0$,
and its Maclaurin expansion has the following form,
\begin{align}
  g_{n}(\omega)=g_{n}(0)-C_{n}\omega^{2n}+\cdots,
  \label{eq:maclaurin}
\end{align}
where $C_{n}=n/(\Gamma(1/(2n))\Delta^{2n+1})$ is positive.
We remark that the generalized Lorentzian distribution introduced in \cite{pietras2018} also has the same expansion form
up to the second leading term.
In the limit $n\to\infty$, $g_{n}(\omega)$ converges to $g_{\infty}(\omega)$ in the $L^{1}$-norm,
\begin{align}
  g_{\infty}(\omega)=\left\{
  \begin{array}{ll}
    1/(2\Delta), & \omega\in(-\Delta,\Delta),\\
    0, & \mathrm{otherwise}.
  \end{array}
  \right.
\end{align}
This distribution is a uniform distribution on a compact support.
See Fig.~\ref{fig:g_n} for the graphs of the distributions $g_{n}(\omega)$
and convergence to $g_{\infty}(\omega)$.

To visualize the extent of synchronization of oscillators,
we introduce the order parameter $r_{N}$ defined by
\begin{equation}
  \label{eq:order-parameter}
  r_{N}=\left|\frac{1}{N}\sum_{j=1}^{N}e^{i\theta_{j}}\right|.
\end{equation}
The order parameter represents the centroid of the oscillators moving on the complex unit circle $\mathbb{S}^{1}$.
When the oscillators are uniformly distributed on $\mathbb{S}^{1}$,
which corresponds to the non-synchronized state, $r_{N}$ gets close to 0.
On the other hand, when the oscillators gather at a point on $\mathbb{S}^{1}$,
which corresponds to the synchronized state, $r_{N}$ equals to 1.
The order parameter $r_{N}$ is therefore useful for monitoring synchronization of the coupled phase-oscillator models.
In the next section,
we will look into the dependency of the order parameter $r_{N}$
on the coupling strength $K$.


The coupled phase-oscillator model on the small-world network
represented by Eq.~\eqref{eq:sw-model} has been considered previously
\cite{chiba2018,medvedev2014},
but we stress that the numbers of links are completely different from ours.
From the construction algorithm, a small-world network has $kN$ links
and $k=3$ in our networks while $k=O(N)$ in the literature.
An advantage of networks with $k=O(N)$ is that they can be analyzed
through the equation of continuity \cite{laszlo2012}.
Nevertheless, this advantage implies at the same time
that such a small-world network in the literature
is essentially the same with the all-to-all coupling,
and is not suitable for detecting new universality classes.


\section{Numerical simulations}
\label{sec:nu-sim}
In the large population limit $N\to\infty$,
the coupled phase-oscillator model, Eq.~(\ref{eq:sw-model}), is
expected to show a synchronization transition around a critical point $K_{\mathrm{c}}$.
For $K<K_{\mathrm{c}}$, the order parameter $r(K):=\lim_{N\to\infty}r_{N}(K)$ is zero,
which corresponds to the non-synchronized state.
On the other hand, for $K>K_{\mathrm{c}}$,
the model shows partially synchronized states, in which $r(K)$ exhibits power law behavior close to the critical point
in the form of
\begin{align}
  r(K)\sim(K-K_{\mathrm{c}})^{\beta},
\end{align}
where $\beta$ is one of the critical exponents.
The critical exponents are crucial to describe critical phenomena,
and models are classified into universality classes, each of which shares the same critical exponents.
Calculating the critical exponents, including $\beta$, is therefore
an important topic from theoretical and numerical perspectives.

\subsection{Finite-size scaling}
The critical exponent $\beta$ is defined in the large population limit
$N\to\infty$, but the limit cannot be achieved through the numerical
simulations. To overcome this difficulty, we use the finite-size
scaling theory, which provides us the limit from observations
in finite-size systems.
The first assumption of our finite-size scaling theory is
existence of the coherent number $N_{\rm c}(K)$ \cite{botet1982} diverging at
the critical point $K=K_{\rm c}$ as
\begin{equation}
  N_{\rm c}(K) \propto (K-K_{\rm c})^{-\bar{\nu}},
\end{equation}
where $\bar{\nu}$ is another unknown positive critical exponent.
The coherent number corresponds to the correlation length
in a simple lattice model.
The second assumption is that the order parameter $r_{N}(K)$
depends on $K$ only through the ratio
\begin{equation}
  \dfrac{N}{N_{\rm c}(K)} \propto \left[ (K-K_{\rm c})N^{1/\bar{\nu}} \right]^{\bar{\nu}}.
\end{equation}
These assumptions imply that $r_{N}(K)$ can be represented by
\begin{equation}
  \label{eq:finite-size}
  r_{N}(K) = N^{-\beta/\bar{\nu}} F((K-K_{\rm c})N^{1/\bar{\nu}}),
\end{equation}
where the function $F$, which is called the scaling function, must be
\begin{equation}
  F(x) \propto x^{\beta} \quad \text{for large } x
\end{equation}
to reproduce the critical exponent $\beta$ in the limit $N\to\infty$.
We remark that the exponent $\beta/\bar{\nu}$ expresses
the finite-size fluctuation of $r_{N}(K)$
at the critical point $K=K_{\rm c}$.

The finite-size scaling is widely used for numerical studies
of critical phenomena in continuous phase transitions,
including coupled phase-oscillator models \cite{hong2002,pelisseto2002,hasenbusch2010,pikovsky2015,hong2015,coletta2017,juhasz2019}.
An important remark on Eq.~\eqref{eq:finite-size} is that,
on the $((K-K_{\rm c})N^{1/\bar{\nu}}, N^{\beta/\bar{\nu}}r_{N})$ plane,
observed values of $r_{N}(K)$ must collapse on a single graph of $F$
for any values of $N$ and $K$.
The unknown values of $K_{\rm c}, \beta$, and $\bar{\nu}$
are determined by detecting the best fit values.
The detection will be performed by using the Bayesian scaling analysis
\cite{harada2011,harada2015},
whose brief introduction is given in Appendix \ref{sec:bsa}.

\subsection{Computation of the order parameter}

We determine the value of the order parameter $r_{N}(K)$
for a given set of $(N,K)$
through temporal evolution of the system and
two steps of averaging.
The model equation, Eq.~\eqref{eq:sw-model},
is numerically integrated by using the fourth-order Runge--Kutta algorithm
with the time step $\delta t=0.1$.
Initial values of the phases $\{\theta_{i}\}$
are randomly drawn from the uniform distribution on the interval $[0,2\pi)$,
and the natural frequencies $\{\omega_{i}\}$ are randomly drawn
from the distribution function $g_{n}(\omega)$.
The order parameter $r_{N}$ defined by Eq.~\eqref{eq:order-parameter}
depends on time $t$,
and we take the time average in the time interval $t\in [300,500]$.
This is the first averaging.

Further, we perform $400$ realizations by changing small-world networks,
the initial values of $\{\theta_{i}\}$, and $\{\omega_{i}\}$
for a given set of $(N,K)$.
To compute the confidence interval of the order parameter,
the resampling technique is in use.
We choose $200$ samples out of $400$ realizations,
and calculate the mean of the time-averaged order parameter
in the chosen $200$ samples.
The mean of the $i$th resampling is denoted by $r_{N}^{(i)}(K)$,
and we perform the resampling for $S=1000$ times.
The value $r_{N}(K)$ is determined by taking the second averaging
over $S$ samples $\{r_{N}^{(i)}(K)\}_{i=1}^{S}$,
which also provide the confidence interval of $r_{N}(K)$.

\begin{figure}[t]
  \begin{center}
    \includegraphics[width=8cm]{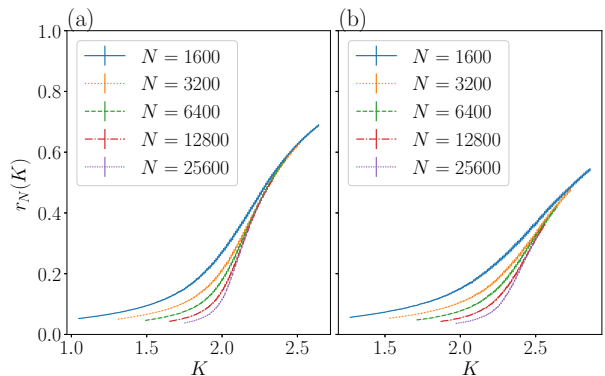}
  \end{center}
  \caption{
    Graphs of order parameter $r_{N}(K)$ with its confidence interval for the model \eqref{eq:sw-model},
    where we take the coupling function $f_{a}(\theta)$
    with (a) $a=0$ and (b) $a=-0.2$.
    As a natural frequency distribution,
    we use $g_{1}(\omega)$ with $\Delta=1$,
    and $N=1600, 3200, 6400, 12800,$ and $25600$ from top to bottom.
    $r_{N}(K)$ and its confidence interval are evaluated by the resampling technique.
    Errorbars are so small that they may not be visible.
  }
  \label{fig:bif-fss}
\end{figure}

See Fig.~\ref{fig:bif-fss} for the obtained $r_{N}(K)$ for $a=0$ and $-0.2$ with $n=1$,
where the condition $a\leq 0$ is expected to give a continuous transition.
In the following two sections, we compute the critical exponents
for $a=0$ and $a=-0.2$,
and show discontinuity for $a=0.5$, respectively.
We remark that $a=-0.2$ and $0.5$ are not special values.
They are arbitrarily chosen from a neighborhood of $a=0$ to demonstrate
differences among the three cases of (i) $a=0$, (ii) $a<0$, and (iii) $a>0$.

\subsection{Critical exponents for continuous transition}

\begin{figure}[t]
  \begin{center}
    \includegraphics[width=8cm]{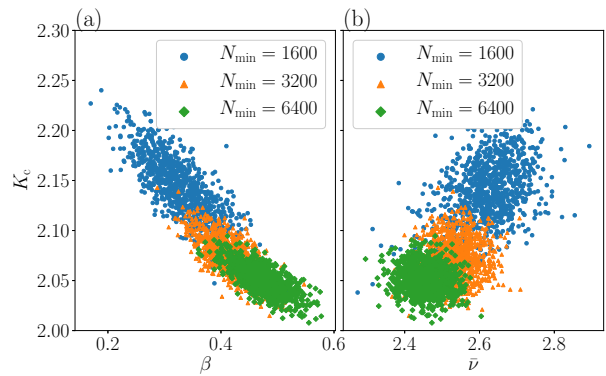}
  \end{center}
  \caption{
    Scattering plots of computed parameters (a)$(\beta,K_{\mathrm{c}})$ and (b)$(\bar{\nu},K_{\mathrm{c}})$, 
    evaluated by the Bayesian scaling analysis.
    Here, we use $(a,n)=(0,1)$, and we set $N_{\min}$ to $1600,3200$ and $6400$.
  }
  \label{fig:scatter}
\end{figure}

\begin{figure}[t]
  \begin{center}
    \includegraphics[width=8cm]{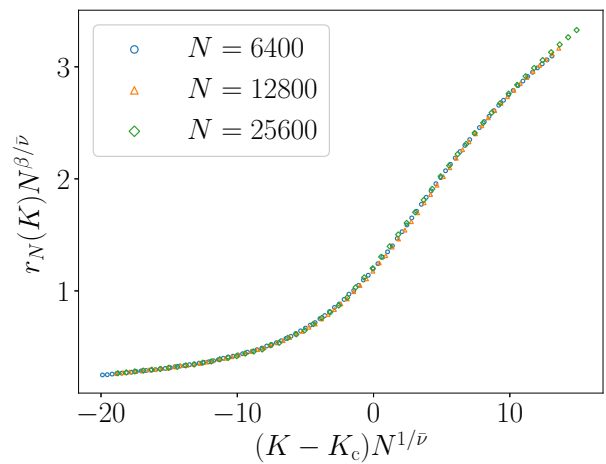}
  \end{center}
  \caption{
    Graph of scaled order parameter $r_{N}(K)N^{\beta/\bar{\nu}}$ versus scaled coupling constant $(K-K_{\mathrm{c}})N^{1/\bar{\nu}}$
    for $(a,n)=(0,1)$,
    where we use $\beta,\bar{\nu}$ and $K_{\mathrm{c}}$, obtained by the Bayesian scaling analysis for $N_{\min}=6400$.
    The values of $\beta,\ \bar{\nu}$, and $K_{\mathrm{c}}$ are shown in Table \ref{table:parameter-estimate}.
    We see that the scaled data are well collapsed to the scaling function $F$.
  }
  \label{fig:fit}
\end{figure}

The finite-size scaling, Eq.~\eqref{eq:finite-size}, is
a powerful tool to compute the unknown values of
$K_{\rm c},\beta$, and $\bar{\nu}$,
but it is not perfect if $N$ is not sufficiently large.
We thus compute the unknown variables
for three values of $N\in\{N_{\rm min},~ 2N_{\rm min}, ~4N_{\rm min}\}$,
and observe convergence by varying $N_{\rm min}$.
Moreover, we use the resampling technique again
to estimate the unknown values with their confidence intervals.
Consequently, we have $S=1000$ sets of the three values
for a given $N_{\rm min}$ as reported in Fig.~\ref{fig:scatter}
because each resampling set $r_{N}^{(i)}(K)$ determines them.
Finally, the values and the confidence intervals
of $K_{\rm c},\beta$, and $\bar{\nu}$
are computed as the averages and the standard deviations
over $S=1000$ sets.
The estimated values are verified in Fig.~\ref{fig:fit},
where all the points lie on a single curve representing
the scaling function $F$ for $N_{\rm min}=6400$.

\begin{table*}[t]
  \begin{center}
    \caption{Critical exponents $\beta,\bar{\nu}$ and the critical point $K_{\mathrm{c}}$ of \eqref{eq:sw-model} depending on the coupling function $f_{a}(\theta)=\sin\theta+a\sin2\theta$ and the natural frequency distribution $g_{n}(\omega)$ in \eqref{eq:g_n}, for $a=0$ and $-0.2$ and $n=1,2,3$, and $\infty$.
    For each pair of $(a,n)$, we use $N_{\min}=1600,3200,6400$, and execute the Bayesian scaling analysis~\cite{harada2011} to find the best parameters fitting \eqref{eq:finite-size}.
    We extrapolate the critical values to $N_{\min}=\infty$ by using the least square method,
    and they are listed in the line of $N_{\min}=\infty$.
    Here, we show the confidence intervals for the last digit of the estimated values 
    in parentheses; for example, $2.13(3)=2.13\pm0.03$.
  }
    \label{table:parameter-estimate}
    \begin{tabular*}{\linewidth}{@{\extracolsep{\fill}}cccccc}\hline
      $f_{a}(\theta)$ & $g_{n}(\omega)$ & $N_{\min}$ & $K_{\mathrm{c}}$ & $\beta$ & $\bar{\nu}$ \\\hline\hline
      $a=0$ & $n=1$ & $1600$ & $2.13(3)$ & $0.33(4)$ & $2.61(7)$ \\
       & & $3200$ & $2.07(1)$ & $0.42(3)$ & $2.53(5)$ \\
       & & $6400$ & $2.05(1)$ & $0.47(3)$ & $2.45(4)$ \\
       & & $\mathbf{\infty}$ & $\mathbf{2.02(2)}$ & $\mathbf{0.51(4)}$ & $\mathbf{2.40(6)}$ \\\cline{2-6}
       & $n=2$ & $1600$ & $1.85(1)$ & $0.27(2)$ & $2.67(5)$\\
       & & $3200$ & $1.78(1)$ & $0.37(2)$ & $2.53(3)$\\
       & & $6400$ & $1.755(9)$ & $0.44(2)$ & $2.50(3)$\\
       & & $\mathbf{\infty}$ & $\mathbf{1.72(1)}$ & $\mathbf{0.49(2)}$ & $\mathbf{2.43(4)}$ \\\cline{2-6}
       & $n=3$ & $1600$ & $1.80(1)$ & $0.28(2)$ & $2.62(4)$\\
       & & $3200$ & $1.76(1)$ & $0.33(2)$ & $2.51(3)$\\
       & & $6400$ & $1.723(8)$ & $0.44(2)$ & $2.51(3)$\\
       & & $\mathbf{\infty}$ & $\mathbf{1.69(1)}$ & $\mathbf{0.47(2)}$ & $\mathbf{2.46(4)}$ \\\cline{2-6}
       & $n=\infty$ & $1600$ & $1.83(1)$ & $0.27(1)$ & $2.50(4)$ \\
       & & $3200$ & $1.79(1)$ & $0.36(2)$ & $2.52(3)$ \\
       & & $6400$ & $1.780(8)$ & $0.41(2)$ & $2.46(3)$ \\
       & & $\mathbf{\infty}$ & $\mathbf{1.76(1)}$ & $\mathbf{0.46(2)}$ & $\mathbf{2.46(4)}$ \\\hline
       $a=-0.2$ & $n=1$ & $1600$ & $2.43(5)$ & $0.38(8)$ & $2.67(9)$ \\
       & & $3200$ & $2.35(2)$ & $0.44(6)$ & $2.58(7)$ \\
       & & $6400$ & $2.34(1)$ & $0.45(4)$ & $2.42(6)$ \\
       & & $\mathbf{\infty}$ & $\mathbf{2.31(3)}$ & $\mathbf{0.48(6)}$ & $\mathbf{2.36(8)}$ \\\cline{2-6}
       & $n=2$ & $1600$ & $2.09(3)$ & $0.31(4)$ & $2.87(7)$\\
       & & $3200$ & $1.99(2)$ & $0.41(4)$ & $2.65(5)$\\
       & & $6400$ & $1.96(1)$ & $0.47(3)$ & $2.52(4)$\\
       & & $\mathbf{\infty}$ & $\mathbf{1.91(2)}$ & $\mathbf{0.51(4)}$ & $\mathbf{2.41(6)}$ \\\cline{2-6}
       & $n=3$ & $1600$ & $2.04(3)$ & $0.27(5)$ & $2.85(8)$\\
       & & $3200$ & $1.96(2)$ & $0.37(4)$ & $2.65(5)$\\
       & & $6400$ & $1.91(1)$ & $0.49(3)$ & $2.59(4)$\\
       & & $\mathbf{\infty}$ & $\mathbf{1.86(2)}$ & $\mathbf{0.55(4)}$ & $\mathbf{2.50(6)}$ \\\cline{2-6}
       & $n=\infty$ & $1600$ & $2.08(3)$ & $0.25(4)$ & $2.76(6)$ \\
       & & $3200$ & $2.00(1)$ & $0.38(4)$ & $2.69(5)$ \\
       & & $6400$ & $1.97(1)$ & $0.43(3)$ & $2.54(4)$ \\
       & & $\mathbf{\infty}$ & $\mathbf{1.94(2)}$ & $\mathbf{0.49(4)}$ & $\mathbf{2.49(6)}$ \\\hline
    \end{tabular*}
  \end{center}
\end{table*}

\begin{figure}[t]
  \begin{center}
    \includegraphics[width=8cm]{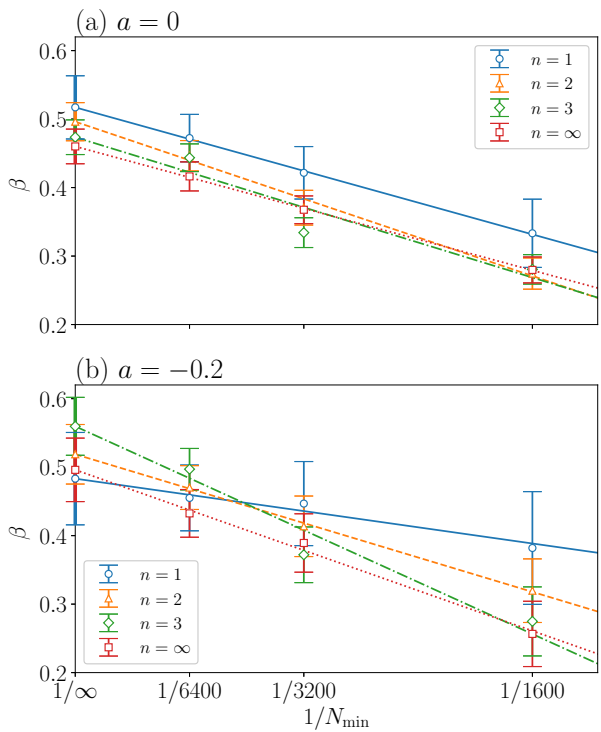}
  \end{center}
  \caption{Graphs of $\beta$ as a function of $1/N_{\min}$
    for (a) $a=0$ and (b) $a=-0.2$ in Eq.~\eqref{eq:sw-model}.
    Critical exponents obtained by the finite-size scaling
    are shown with errorbars,
    and the least square method gives the extrapolations
    at the left boundary of the panels.
    For each $a$, the resulting linear regression lines are drawn with the solid line for $n=1$,
    the dashed line for $n=2$, the dot-dashed line for $n=3$, and the dotted line for $n=\infty$.
  }
  \label{fig:ce_extrapolation}
\end{figure}

The estimated values of $K_{\rm c}, \beta$ and $\bar{\nu}$
are summarized in Table~\ref{table:parameter-estimate}.
The row of $N_{\min}=\infty$ is obtained by extrapolation
from $N_{\min}=1600, 3200$, and $6400$
as demonstrated in Fig.~\ref{fig:ce_extrapolation}.
We note that the extrapolated values of $\beta$ are close to $1/2$
and ones of $\bar{\nu}$ are close to $5/2$
irrespective of the values of $a$ and $n$.
The universality is completely unlike the all-to-all interaction case.
Here we note that this result shares the same critical exponent $\bar{\nu}=5/2$
as the all-to-all interaction case for $(a,n)=(0,1)$
calculated in \cite{hong2007}.

The value $\bar{\nu}\simeq 5/2$ is not in agreement
with the value $\bar{\nu}\simeq 2$ previously reported
for $(a,n)=(0,1)$ \cite{hong2002}.
We suppose that the discrepancy comes from the method
to compute the critical exponents.
In the literature, the authors used the fact
that $r_{N}(K)N^{\beta/\bar{\nu}}$ takes a constant value
irrespective of $N$ at the critical point $K=K_{\rm c}$
(See Eq.~\eqref{eq:finite-size}). Using this fact,
they first find the best fit values of $\beta/\bar{\nu}$ and $K_{\rm c}$
by varying the system size $N$. One more equation is obtained
by derivating the finite-size scaling, Eq.~\eqref{eq:finite-size},
which produces
\begin{equation}
  \log\left[ \frac{\diff r_{N}}{\diff K}(K_{\mathrm{c}}) \right]
  = \dfrac{1-\beta}{\bar{\nu}} \log N + \mathrm{const.}
\end{equation}
Plotting the left-hand side as a function of $\log N$,
one has the slope $(1-\beta)/\bar{\nu}$.
A remarkable disadvantage of this method is that
the estimation relies on high precision of $r_{N}(K)$
around the critical point $K=K_{\rm c}$,
while the Bayesian scaling analysis uses $r_{N}(K)$
in a wider interval of $(K-K_{\rm c})N^{1/\bar{\nu}}$
and provides persistence against fluctuation.
We, therefore, believe that $\bar{\nu}\simeq 5/2$
obtained by the Bayesian scaling analysis is more reliable.

\subsection{Discontinuity of transition}

\begin{figure}[t]
  \begin{center}
    \includegraphics[width=8cm]{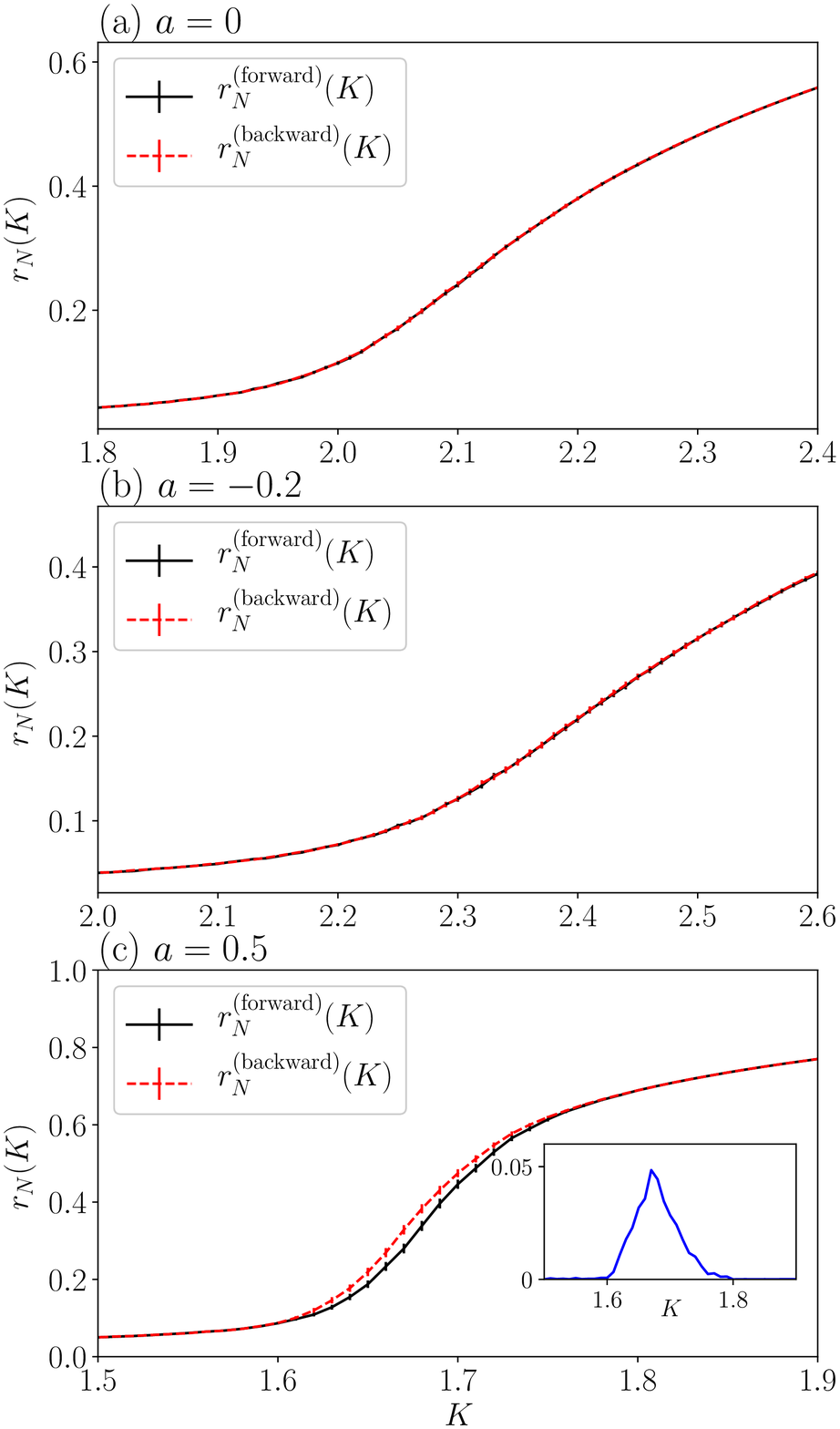}
    \caption{
      Graphs of $r_{N}(K)$ and its errorbar of \eqref{eq:sw-model}
      for (a) $(a,n)=(0,1)$, (b) $(a,n)=(-0.2,1)$, and (c) $(a,n)=(0.5,1)$
      with two different types of initial phases,
      where we set the number of oscillators $N=25600$.
      We see that, only in (c), $r_{N}(K)$ takes a different value depending on the choice of the initial phases
      around $K\in(1.6,1.8)$.
      The inset in (c) shows the graph of $r_{N}^{(\mathrm{backward})}(K)-r_{N}^{(\mathrm{forward})}(K)$.
    }
    \label{fig:hysteresis}
  \end{center}
\end{figure}

In the all-to-all interaction
a positive $a$ induces discontinuity of the synchronization transition
\cite{chiba2011}.
We reveal that the transition is discontinuous
also in a small-world network.
The discontinuity appears as a result of a subcritical transition,
and a subcritical transition has metastability:
A partially synchronized state is stable in addition to
a stable nonsynchronized state for a fixed $K$ close to the critical point.
The metastability implies that the final state depends on
choice of the initial state,
and the dependency is extracted by observing hysteresis.

Fixing $a=0.5$, we check existence of the hysteresis
by preparing two sets of the initial phases $\{\theta_{i}\}_{i=1}^{N}$
for each $K$:
(i) We start from $K=K_{\mathrm{start}}$,
where $K_{\rm start}$ is sufficiently smaller than the critical value $K_{\rm c}$,
and the initial phases $\{\theta_{i}\}_{i=1}^{N}$ are randomly drawn
from the interval $[0,2\pi)$.
At a certain value of $K$, the final phases at $t=500$ is used
as the initial phases at the successive value $K+\Delta K$
in the increasing direction.
The increase of $K$ is continued up to $K=K_{\rm end}$,
where $K_{\rm end}$ is sufficiently larger than the critical value $K_{\rm c}$.
%
We call the process (i) the ``forward'' process, and $r_{N}^{(\mathrm{forward})}(K)$ denotes its order parameter.
(ii) Contrary to the ``forward'' process,
we start with the random initial phases $\{\theta_{i}\}_{i=1}^{N}$
at $K=K_{\mathrm{end}}$
and decrease $K$ up to $K=K_{\rm start}$
following the same procedure with the ``forward'' process.
We call this process the ``backward'' process, and $r_{N}^{(\mathrm{backward})}(K)$ denotes its order parameter.
We have executed the numerical simulations of Eq.~\eqref{eq:sw-model}
for $a=0,-0.2$, and $0.5$, and $n=1,2,3$, and $\infty$.
For the system size $N=25600$,
the hysteresis appears only for $a=0.5$ regardless of $n$
as exampled in Fig.~\ref{fig:hysteresis} for $n=1$.
We have checked that $t=500$ is sufficiently long
to pass the transient period,
and simulations up to $t=800$ do not affect the hysteresis.
We therefore conclude that the system represented by Eq.~\eqref{eq:sw-model}
shows a discontinuous transition for $a=0.5$ as the all-to-all interaction case.

\section{Small-world network and noise}
\label{sec:SW-noise}

We discuss similarity between systems on small-world networks
and noise systems.
For simplicity, we consider the Kuramoto model $(a=0)$ for a while.
The steady state in the Kuramoto model is
proportional to $\delta(\omega-Kr\sin\theta)$
in the synchronized regime
of $\omega$ \cite{strogatz2000,fonseca2018},
where $\delta$ is the Dirac's delta function.
The $\delta$ function with the integration over $\omega$
and symmetry of the natural frequency distribution
yield the self-consistent equation of the order parameter $r$ as
\begin{align}
  r=Kr\int_{-\pi/2}^{\pi/2}g_{n}(Kr\sin\theta)\cos^{2}\theta\diff\theta.
  \label{eq:self-consistent}
\end{align}
The order parameter $r$ is sufficiently small around the critical point
and we perform the Maclaurin expansion of $g_{n}$.
The leading order of the expansion, which is of $O(r)$,
determines the celebrated critical point $K_{\rm c}=2/[\pi g_{n}(0)]$.
The partially synchronized branch is obtained
by balancing the second leading order of $O(r^{2n+1})$
with the first leading order of $O(r(K-K_{\rm c}))$,
and the balance results to $r\propto (K-K_{\rm c})^{1/(2n)}$.
We then obtain the critical exponent $\beta=1/(2n)$.

\begin{figure}[t]
    \centering
    \includegraphics[width=8cm]{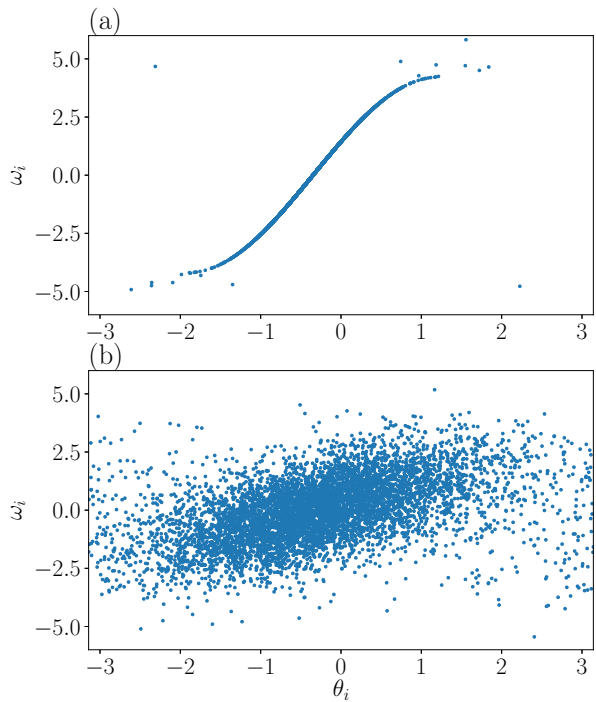}
    \caption{
      Snap shots of oscillators on the $(\theta_{i},\omega_{i})$ plane
      at $t=500$.
      (a) The all-to-all network.  (b) A small-world network.
      The system size $N=6400$. The coupling constant $K=4.5$.
      $(a,n)=(0,1)$.
    }
    \label{fig:all-sw-scatter}
\end{figure}

To the contrary, on a small-world network,
a steady state is not written in the form of the $\delta$ function
and the synchronized oscillators are still ``noisy''
as shown in Fig.~\ref{fig:all-sw-scatter}.
The synchronized oscillators no longer capture
the flatness of $g_{n}(\omega)$ around $\omega=0$,
and the critical exponent $\beta$ falls into the classical value $1/2$
regardless of natural frequency distribution $g_{n}(\omega)$
as a noisy system \cite{sakaguchi1988}.

Moreover, in the model having the nonvanishing second harmonics
of the coupling function with $a<0$,
the noise recovers $\beta=1/2$ \cite{crawford1995}
whereas no noise system gives $\beta=1$ \cite{daido1994}.
The universality of $\beta=1/2$ observed in systems on small-world networks
is therefore very similar to the one in noisy systems.

\section{Conclusion and Discussion}
\label{sec:conclusion}
We calculated the critical exponents $\beta$ and $\bar{\nu}$
for coupled phase-oscillator models on small-world networks
by using the finite-size scaling method.
We set the coupling function as $f_{a}(\theta)=\sin\theta+a\sin2\theta$, and the natural frequency distribution
as $g_{n}(\omega)$ defined in Eq.~\eqref{eq:g_n},
and we studied the $(a,n)$-dependency of the critical exponents.
Our numerical results suggest $\beta=1/2$ and $\bar{\nu}=5/2$
for all $g_{n}(\omega)$ and coupling function $f_{a}(\theta)$ with $a=0$ and $-0.2$.
This universality shows a sharp contrast with the all-to-all interaction case,
which has various values of $\beta$ depending on the coupling function and the natural frequency distribution.
A possible explanation of the source of contrast
can be found in the number of links of considering networks:
our small-world networks has $O(N)$ links,
while the all-to-all interaction have $O(N^{2})$ links.
We have also found that the model, Eq.~\eqref{eq:sw-model},
shows a discontinuous transition for $a=0.5$.
The (dis)continuity is a weaker property than the values of the critical exponents,
and it is shared between the two types of networks:
networks with $O(N)$ links and $O(N^{2})$ links.


We end this paper commenting on two future works.
Firstly,
we picked up two representative points of $a$ from a neighborhood of $a=0$
to investigate universality of the critical exponents.
Studying a global phase diagram on the $(K,a)$-plane is a subject for future researches.
Secondly, we note universal value $\beta=1/2$ in the Kuramoto model
which is recovered by adding noise regardless of the natural
frequency distribution \cite{sakaguchi1988}.
A small-world network may play a role of noise due to
inhomogeneous couplings,
and another work to do is to make a bridge between a noisy Kuramoto model
and a model on a small-world network.

\acknowledgments
In this research work we used the supercomputer of ACCMS, Kyoto University.
R.~Y. acknowledge the support of Iwadare Scholarship from Iwadare Scholarship Foundation.
Y.~Y.~Y. acknowledges the support of JSPS KAKENHI Grant No. 16K05472.

\appendix
\section{Bayesian Scaling Analysis}
\label{sec:bsa}
We briefly review the Bayesian scaling analysis~\cite{harada2011,harada2015},
 a statistical method for estimating the values such as $\beta, \bar{\nu},$ and $K_{\rm c}$ in Eq.~\eqref{eq:finite-size}.
We write these values as $\theta_{p}=(\beta,\bar{\nu},K_{\mathrm{c}})$.
We assume that the scaling function $F$ in Eq.~\eqref{eq:finite-size} obeys a Gaussian process
\begin{align}
    F\sim\mathcal{GP}(m,k_{\theta_{h}}),
\end{align}
with mean function $m(\cdot)$ and covariance kernel $k_{\theta_{h}}(\cdot,\cdot)$.
Here $\theta_{h}$ denotes the hyperparameters of covariance kernel.
We also set $m=0$ for simplicity.
In the following, we also use the notation $\bm{\theta}=(\theta_{h},\theta_{p})$.
For the data $\{r_{N_{i}}(K_{i})\}_{i=1}^{M}$,
the rescaled data $X_{\theta_{p},i}=(K_{i}-K_{\mathrm{c}})N_{i}^{1/\bar{\nu}}$
and $Y_{\theta_{p},i}=r_{N_{i}}(K_{i})N_{i}^{\beta/\bar{\nu}}$
must collapse on the scaling function as $Y_{\theta_{p},i}=F(X_{\theta_{p},i})$.
Since $F$ is a Gaussian process, $Y_{\theta_{p}}$ obeys a $M$-dimensional
Gaussian distribution, and the probability of $Y$ for the parameter $\bm{\theta}$ is
\begin{align}
    &p(Y\mid\bm{\theta})=\mathcal{N}(Y_{\theta_{p}}\mid\bm{0},K_{\bm{\theta}})\notag\\
    =&\frac{1}{(2\pi)^{N/2}[\det K_{\bm{\theta}}]^{1/2}}
    \exp\left[-\frac{1}{2}Y_{\theta_{p}}^{\mathsf{T}}K_{\bm{\theta}}^{-1}Y_{\theta_{p}}\right].
    \label{eq:posterior}
\end{align}
Here, $[K_{\bm{\theta}}]=k_{\theta_{h}}(X_{\theta_{p},i},X_{\theta_{p},j})$ is $M\times M$ dimensional matrix.
By assuming that the prior distribution of $\bm{\theta}$ is uniform, we have
\begin{align}
    p(\bm{\theta}\mid Y)\propto p(Y\mid\bm{\theta}),
\end{align}
from Bayes' theorem.
The most probable parameters $\bm{\theta}$ are, therefore,  estimated by finding the minimum of likelihood function given by
\begin{align}
  L_{\bm{\theta}}
  = \log(\det K_{\bm{\theta}})
  + Y_{\theta_{p}}^{\mathsf{T}}K_{\bm{\theta}}^{-1}Y_{\theta_{p}},
    \label{eq:likelihood}
\end{align}
which is obtained by taking $\log$ and discarding constants in Eq.~\eqref{eq:posterior}.
The gradient of $L_{\bm{\theta}}$ for an element $\theta\in\bm{\theta}$ is given by
\begin{equation}
\begin{split}
    \frac{\partial L_{\bm{\theta}}}{\partial \theta}
    =&\mathrm{tr}\left[K_{\bm{\theta}}^{-1}\frac{\partial K_{\bm{\theta}}}{\partial\theta}\right]-(K_{\bm{\theta}}^{-1}Y_{\theta_{p}})^{\mathsf{T}}\frac{\partial K_{\bm{\theta}}}{\partial \theta}(K_{\bm{\theta}}^{-1}Y_{\theta_{p}})\\
    &+2Y_{\theta_{p}}^{\mathsf{T}}K_{\bm{\theta}}^{-1}\frac{\partial Y_{\theta_{p}}}{\partial\theta},
\end{split}
\end{equation}
and using this gradient, the gradient method gives us the most probable parameters $\bm{\theta}$.

In this paper, we consider a kernel based on a radial basis function (RBF) kernel
\begin{align}
    k_{\theta_{h}}(x,y)=\theta_{1}\exp\left[-\frac{(x-y)^{2}}{\theta_{2}}\right]+\theta_{3}\delta(x,y),
\end{align}
which is parameterized by $\theta_{h}=(\theta_1,\theta_2,\theta_3)$ with $\theta_{1,2,3}>0$,
and $\delta(x,y)=1$ when $x=y$, otherwise $\delta(x,y)=0$.
Here, $\theta_{3}$ denotes the data fidelity.
Roughly speaking, a sample path of Gaussian process associated with a RBF kernel are known to be an infinitely differentiable function;
see \cite[Corollary~4.13]{kanagawa2018} for a rigorous statement.
Therefore, the Bayesian scaling analysis only assumes the smoothness of a scaling function,
and it does not need an explicit form.
See the reference \cite{harada2011,harada2015} for more detailed discussions.

\end{document}